\documentclass[12pt]{article}
\usepackage{epsfig}
\usepackage{amssymb}

\def\b{\beta}

\def\f{\phi}
\def\g{\gamma}
\def\h{\eta}

\def\k{\kappa}

\def\m{\mu}
\def\n{\nu}

\def\p{\pi}

\def\s{\sigma}
\def\t{\tau}

\def\D{\Delta}

\def\G{\Gamma}

\def\L{\Lambda}

\def\ve{\varepsilon}

\def\mt{\widetilde{m}_1}                  
\def\mb{\overline{m}}
\def\beq{\begin{equation}}
\def\eeq{\end{equation}}

\def\bea{\begin{eqnarray}}
\def\eea{\end{eqnarray}}

\def\pl#1#2#3{Phys.~Lett.~{\bf B {#1}} ({#2}) #3}
\def\np#1#2#3{Nucl.~Phys.~{\bf B {#1}} ({#2}) #3}
\def\prl#1#2#3{Phys.~Rev.~Lett.~{\bf #1} ({#2}) #3}

\def\ap#1#2#3{Ann.~of Phys.~{\bf {#1}} ({#2}) #3}

\begin{document}
\date{}

\title{ 
{\normalsize
\mbox{ }\hfill
\begin{minipage}{3cm}   
DESY 02-107\\
OUTP-02-36-P
\end{minipage}}\\
\vspace{2cm}
\bf A Bound on Neutrino Masses\\ From Baryogenesis}
\author{W.~Buchm\"uller, P. Di Bari\\
{\it Deutsches Elektronen-Synchrotron DESY, 22603 Hamburg, Germany}\\[5ex]
M.~Pl\"umacher\\
{\it Theoretical Physics, University of Oxford, 1 Keble Road,}\\
{\it Oxford, OX1 3NP, United Kingdom}
}
\maketitle

\thispagestyle{empty}


\begin{abstract}
\noindent
Properties of neutrinos, the lightest of all elementary particles, may be the origin
of the entire matter-antimatter asymmetry of the universe.
This requires that neutrinos are Majorana particles, which are equal to their 
antiparticles, and that their masses are sufficiently small. Leptogenesis, the
theory explaining the cosmic matter-antimatter asymmetry, predicts that
all neutrino masses are smaller  than $0.2$~eV, which will be tested by
forthcoming laboratory experiments and by cosmology. 
\end{abstract}

\newpage

Neutrinos are unique elementary particles. They have only weak interactions and can
therefore travel astronomically large distances without being absorbed. In particular,
they reach the earth from the center of the sun. The study of solar and atmospheric
neutrinos, produced by cosmic rays in the earth's atmosphere, has provided evidence
for neutrino oscillations. These take place between the three species of neutrinos,
$\n_e$, $\n_\m$ and $\n_\t$, which belong to the three families of elementary particles
with the charged leptons electron ($e$), muon ($\m$) and tau ($\t$), respectively.

Neutrinos are electrically neutral, spin-${1\over 2}$ particles.
In the standard model of strong and electroweak interactions neutrinos are massless.
As a consequence, electron-, muon-, and tau-number are conserved. Due to the evidence
for neutrino oscillations we now know that neutrinos have mass and that the three lepton 
numbers are not separately conserved. However, we still do not know whether the total
lepton number, $L=L_e+L_\m+L_\t$, is also violated. In this case neutrinos would be
Majorana particles which are equal to their antiparticles.

The standard model of particle physics can be derived as low energy effective theory
from a more fundamental grand unified theory (GUT), whose symmetry becomes manifest only
at much shorter distances, or correspondingly larger energies. The simplest GUT which
unites the strong and electroweak forces, and where all quarks and leptons can turn
into each other via gauge interactions, is based on the group SO(10) \cite{gfm75}.
This theory predicts the existence of `sterile' neutrinos $\n_R$, one for each quark-lepton
generation, which have no strong or electroweak interactions, unlike all other known
elementary particles. After the breaking of the GUT group SO(10) to the standard model gauge
group SU(3)$\times$SU(2)$\times$U(1), the sterile neutrinos become massive
Majorana fermions, $N=\n_R + \n_R^c$, where the superscript `c' denotes antiparticles.

Heavy Majorana neutrinos with masses $M$ of order the GUT mass scale,
$\L_{G} \sim 10^{15}$~GeV,
also give rise to small neutrino masses through mixing, which is the so-called seesaw
mechanism \cite{yan79}. The light neutrino
mass eigenstates $\n_1$, $\n_2$ and $\n_3$ are mixtures of $\n_e$, $\n_\m$ and $\n_\t$, and 
again Majorana neutrinos, i.e. identical to their antiparticles, $\n_i = \n_i^c$. 
Assuming for the neutrino masses the ordering $m_1 < m_2 < m_3$, one expects for the 
largest mass $m_3$  as consequence of the seesaw mechanism,
\bea\label{nmass}
m_3 \sim {v^2\over M} \sim 0.01\ {\rm eV}\;,
\eea
where we have chosen for $v$ the mass scale of electroweak symmetry breaking of order
the W-boson mass, $v \simeq 174$~GeV.   

The evidence for neutrino oscillations reported by the Super-Kamiokande \cite{ska00} 
and SNO \cite{sno02} collaborations suggests $\n_\m \rightarrow \n_\t$ and 
$\n_e \rightarrow \n_\m$
oscillations. The analysis yields two neutrino mass differences,
$\D m^2_{atm} = (1.6 - 3.9)\times 10^{-3}\ {\rm eV}^2$ \cite{ska00} and  
$\D m^2_{sol} = (0.2 - 2)\times 10^{-4}\ {\rm eV}^2$ \cite{sno02}. It is very remarkable 
that $\sqrt{\D m^2_{atm}} \sim 0.05$~eV and $\sqrt{\D m^2_{sol}} \sim 0.008$~eV are 
consistent with the estimate (\ref{nmass}). This suggests that the extrapolation to
the GUT mass scale $\L_G \sim 10^{15}$~GeV by means of the seesaw mass relation may indeed
be correct! 

So far, however, we do not know the absolute scale of neutrino masses since oscillation
experiments can only determine differences of mass squares. An upper bound on a
weighted sum of neutrino masses has been obtained from a study of the electron
energy spectrum in tritium $\b$-decay, 
$m_{\n_e} = \left(\sum_i |U_{ei}|^2m_i^2\right)^{1/2} < 2.2$~eV \cite{mai01}. 
Here $U_{ei}$ ($i=1\ldots 3$) are elements of the `leptonic mixing matrix' 
which determine the couplings of an electron to a W-boson and
the neutrino mass eigenstates $\n_i$. An even more stringent bound has been obtained
from neutrinoless double $\b$-decay, which for a nucleus with $Z$ protons and $N$
neutrons corresponds to the transition
$A(Z,N) \rightarrow A(Z+2,N-2) + e^- + e^-$. Note, that this
process violates lepton number! Non-observation of neutrinoless double $\b$-decay 
yields an upper bound on the effective neutrino mass 
$m_{ee} = |\sum_i U_{ei}^2m_i| < 0.38 \k$~eV, where $\k={\cal O}(1)$ represent the
uncertainty of the nuclear matrix element \cite{hei01} . 

Important information on neutrino masses also
comes from cosmology. The requirement that neutrinos do not overclose the
universe yields the limit $\sum_i m_{\n_i} < 40$~eV. A much stronger bound has 
recently been obtained from the 2dF Galaxy Redshift Survey, $\sum_i m_{\n_i} < 1.8$~eV
at 95\% confidence level \cite{elg02}, which is comparable to the bound from tritium 
$\b$-decay.

Baryons, i.e. protons and neutrons, dominate by far the cosmic mass fraction of
visible matter. As far as we know, the universe contains no primordial antimatter.
Hence, the baryon density $n_B$ is equal to the difference of baryon and antibaryon
densities, which is conventionally normalized to the photon density, 
$\h_B = (n_B - n_{\bar{B}})/n_\g$.
The precision of measurements of this baryon asymmetry has significantly improved with 
the observation of the acoustic peaks in the cosmic microwave background radiation (CMB).
The BOOMERanG and DASI experiments have measured the baryon asymmetry with a (1$\s$) 
standard error of $\sim 15\%$ \cite{boo02,das02},
\bea\label{cmb}
\eta_{B0}^{CMB}=(6.0^{+1.1}_{-0.8})\times 10^{-10}\;,
\eea
which is consistent with the result of standard primordial nucleosynthesis (BBN),
obtained from the measurement of relic nuclear abundances \cite{rpp02}. 

At first sight, it may appear very surprizing that the observed cosmological
baryon asymmetry should be related to neutrino properties. The reason is a deep
connection \cite{tho76} between baryon and lepton number in the standard model of 
particle physics which is hidden today, but which becomes visible at very high
temperatures above the scale of electroweak symmetry breaking, 
$T > v \sim 100\ {\rm GeV} \sim 10^{15}\ {\rm K}$. At such high temperatures,
which may have been realized in the early universe, quarks and leptons are rapidly
transformed into each other by so-called sphaleron processes \cite{krs85}, and lepton 
and baryon number are no longer separately conserved. In thermal equilibrium baryon
and lepton asymmetries have a constant ratio determined by the standard model.

A baryon asymmetry can be dynamically generated in an expanding universe if the particle
interactions violate baryon number, charge conjugation ($C$) and joint parity and charge
conjugation ($C\!P$), and if the evolution of the universe leads to a departure from 
thermal equilibrium \cite{sak67}. All these conditions can be fulfilled in decays of 
heavy Majorana neutrinos ($N$) into light leptons ($l$) and Higgs bosons ($\f$). 
$C\!P$ violating, out-of-equilibrium decays $N \rightarrow l\f$ generate naturally 
a lepton asymmetry which, by means of the sphaleron processes, is  partially 
transformed into a baryon asymmetry. This is the simple and elegant leptogenesis
mechanism \cite{fy86}.

The non-equilibrium process of baryogenesis is described by means of Boltzmann
equations. The final baryon asymmetry is the result of a competition between
production processes and washout processes which tend to erase any generated
asymmetry. We shall assume that the dominant contribution
is given by decays of $N_1$, the lightest of the heavy Majorana neutrinos.
This assumption is well justified in the case of a mass hierarchy among the
heavy neutrinos, i.e. $M_1 \ll M_2, M_3$, and it is also known to be a good
approximation, if $M_{2,3}-M_1 = {\cal O}(M_1)$. The case of partial
degeneray, $M_{2,3}-M_1 \ll M_1$, requires a special treatment. 
In a recent detailed study \cite{bdp02} it has been 
shown that the final baryon asymmetry depends on just four parameters: 
the mass $M_1$ of $N_1$, the lightest of the heavy Majorana neutrinos; 
the $C\!P$ asymmetry in $N_1$ decays, 
$\ve_1 = (\G(N_1\rightarrow l\f)-\G(N_1\rightarrow l^c\f^c))/
(\G(N_1\rightarrow l\f)+\G(N_1\rightarrow l^c\f^c))$;
the effective neutrino mass $\mt$, with $m_1 \leq \mt \lesssim m_3$, which 
is a measure of the strength of the coupling of $N_1$ to the thermal bath, and,
finally, the sum of all neutrino masses squared, $\mb^2 = m_1^2 + m_2^2 + m_3^2$,
which controls an important class of washout processes. Together with the two
mass squared differences $\D m^2_{atm}$ and $\D m^2_{sol}$, the sum $\mb^2$ 
determines all neutrino masses.

The CP asymmetry $\ve_1$ satisfies an upper bound, $|\ve_1|<\ve(M_1,\mb)$ 
\cite{di02}. The maximal baryon asymmetry produced by $N_1$ decays can then be
expressed in terms of a maximal $C\!P$ asymmetry $\ve$ and an efficiency factor
$\k_0$ \cite{bdp02},
\bea\label{constraint}
\eta_{B0}^{\rm max}(\mt,M_1,\mb) =
0.96\times 10^{-2}\,\ve(M_1,\mb)\,\k_0(\mt,M_1,\mb) \; ,
\eea
with
\bea\label{cpasym}
\ve(M_1,\mb) = {3\over 16\p}\ {M_1\over v^2}\ {\D m^2_{atm}+\D m^2_{sol}\over m_3}\;.
\eea
Note, that the maximal $C\!P$ asymmetry is independent of the mass hierarchy among
the neutrinos. For a `normal' hierarchy, i.e.
$\D m^2_{atm} = m_3^2 - m_2^2 > m_2^2-m_1^2 = \D m^2_{sol}$,
the individual neutrino masses are given by
\bea
m_3^2 &=& {1\over 3}\left(\mb^2 + 2\D m^2_{atm} + \D m^2_{sol}\right)\;, \\
m_2^2 &=& {1\over 3}\left(\mb^2 - \D m^2_{atm} + \D m^2_{sol}\right)\;, \\
m_1^2 &=& {1\over 3}\left(\mb^2 - \D m^2_{atm} - 2\D m^2_{sol}\right)\;. 
\eea
Inserting the expression for $m_3$ into eq.~(\ref{cpasym}) yields
the maximal asymmetry $\ve$ as function of $M_1$ and $\mb$. Analogous formulae hold
in the case of an `inverse' hierarchy  where 
$\D m^2_{sol} = m_3^2-m_2^2 < m_2^2 - m_1^2=\D m^2_{atm}$. The difference between
`normal' and `inverted' hierarchy is irrelevant for the following analysis.

\begin{figure}[t]
\centerline{\epsfig{file=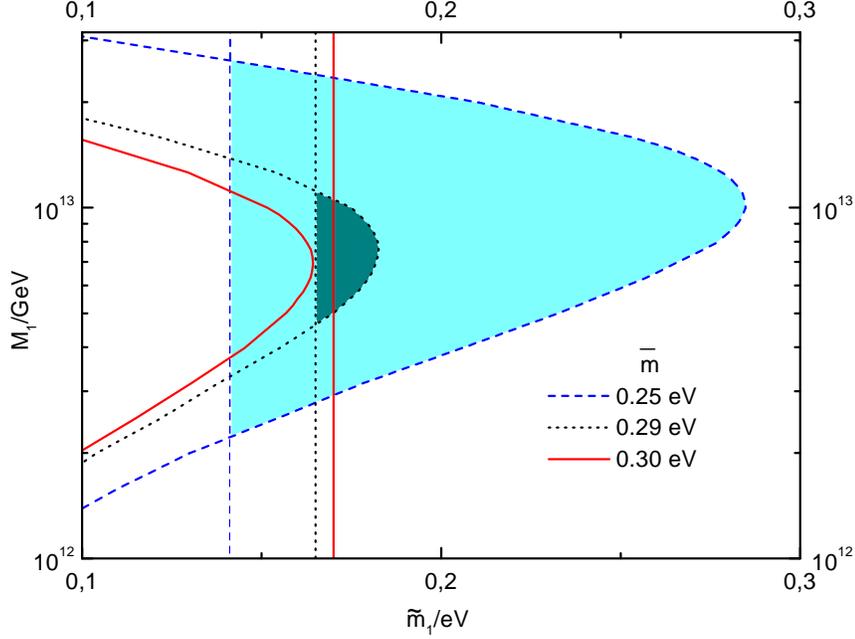,width=14cm}}
\caption{{\it
Regions in the $\mt$-$M_1$-plane allowed by baryogenesis for different values of
$\mb$. The boundaries are defined by  
$\eta_{B0}^{\rm max}(\mt,M_1,\mb)=\eta^{CMB}_{B0}$ and by $\mt=m_1(\mb)$.
}}
\end{figure}

If leptogenesis is the true mechanism of baryogenesis, the maximal baryon asymmetry 
has to be larger than the observed one, for which we choose the most precise CMB 
value (\ref{cmb}), i.e. we require
$\eta_{B0}^{\rm max}(\mt,M_1,\mb)\gtrsim  \eta^{CMB}_{B0}$. 
This condition imposes an important restriction on the space of neutrino parameters 
$\mt$, $M_1$ and $\mb$. Since the strength of washout processes increases with 
increasing $\mb$, one obtains in particular an upper bound on $\mb$. 

\begin{figure}[t]
\centerline{\epsfig{file=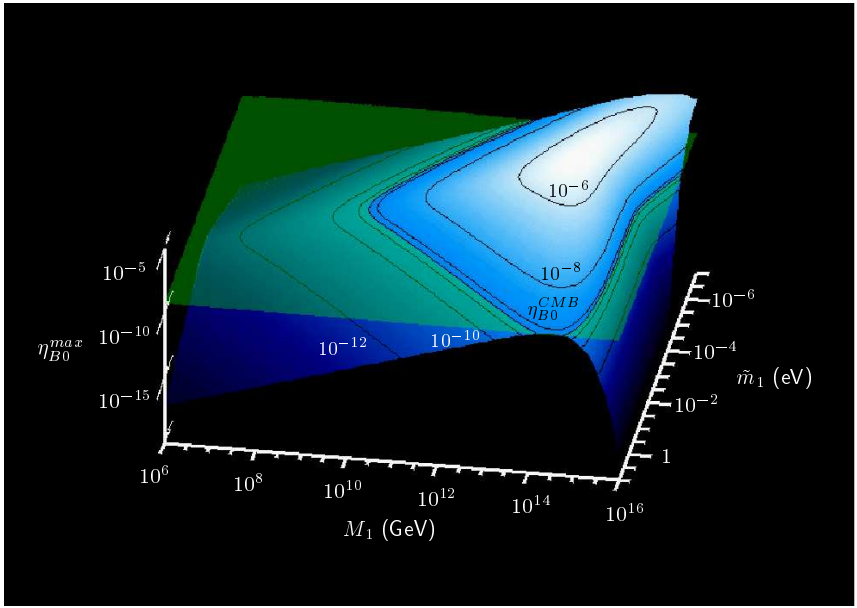,angle=270,width=112mm}}
\caption{{\it 
Maximal baryon asymmetry $\eta_{B0}^{\rm max}$ (blue) as function of $\mt$ and $M_1$ 
for $\mb = 0.05$~eV. The black lines are curves of constant baryon asymmetry with
the value indicated. The lines around the intersection with the green plane 
correspond to the measured value and the upper/lower limits at $3\sigma$.}}
\end{figure}

In order to determine the upper bound on $\mb$ one has to calculate the efficiency 
factor $\k_0$ by solving the Boltzmann equations for different points in the 
neutrino parameter space $(\mt,M_1,\mb)$. Results of a numerical scan are 
displayed in fig.~1 which shows the curves defined by 
$\eta_{B0}^{\rm max}(\mt,M_1,\mb)=\eta^{CMB}_{B0}$ and by $\mt=m_1(\mb)$, 
for $\mb = 0.25$~eV (dashed), $0.29$~eV (dotted) and 
$0.30$~eV (full). For the two smaller values of $\mb$ domains exist which
allow large enough baryon asymmetry. For $\mb = 0.30$~eV successfull baryogenesis
is no longer possible. 

The determination of a theoretical error on the upper bound
$\mb < 0.30$~eV would require a systematic quantitative treatment of baryogenesis,
which goes beyond the Boltzmann equations. This is not yet available. However, all
corrections we are aware of tend to decrease the generated baryon asymmetry. 
Including these corrections would therefore lead to an even stronger upper bound on $\mb$.
Based on the studies in \cite{plu98}, we expect that in the supersymmetric standard
model the bound on $\mb$ is rather similar to the non-supersymmetric case. A detailed
numerical investigation is in progress.

For the upper bound $\mb^{max} = 0.30$~eV the individual neutrino masses are 
$m_3 = 0.18$~eV and $m_1 = m_2 = 0.17$~eV. Using the unitarity of the leptonic
mixing matrix, $\sum_i |U_{ei}|^2 = 1$, one then obtains for the mass measured in
tritum $\b$-decay, $m_{\n_e} < 0.18$~eV. This value is smaller than the sensitivity
aimed at by the KATRIN collaboration \cite{kat01}. The same bound as for $m_{\n_e}$
one also finds for the Majorana  mass measued in neutrinoless double $\b$-decay,
$m_{ee} < 0.18$~eV. This prediction can be tested by the GENIUS project \cite{gen01}.
Cosmological bounds on neutrino masses will also improve, and the Sloan Digital
Sky Survey combined with CMB data from the MAP experiment is expected to reach the 
sensitivity \cite{het98} of the leptogenesis bound, $\sum_i m_i \leq 0.52$~eV.

For strongly hierarchical neutrinos, $m_1 < m_2 \ll m_3 \simeq \mb \simeq 0.05$~eV,
successful baryogenesis is possible for a large domain in the $\mt$-$M_1$-plane.
This can be seen in fig.~2 where the surface $\eta_{B0}^{\rm max}$ is shown
for $\mb = 0.05$~eV. 
The intersection with the plane $\eta_{B0}^{\rm CMB}$ yields the allowed region of
parameters. For the heavy neutrino mass one obtains the lower bound 
$M_1 > 4\times 10^8$~GeV \cite{bdp02}. In many models of neutrino
masses the $C\!P$ asymmetry saturates the upper bound, $|\ve_1| \sim |\ve_1^{max}|$
(cf.~\cite{bdp02}). The CMB baryon asymmetry then yields a precise relation
between $\mt$ and $M_1$. For the particularly interesting range 
$0.001\ {\rm eV} \lesssim \mt \lesssim 0.1\ {\rm eV}$ one has the simple relation
$M_1 \simeq 3\times 10^{10}\ {\rm GeV}\ \mt/(0.01 {\rm eV})$. Such values of $M_1$
are naturally obtained in SO(10) grand unified models where 
lepton number is broken at the GUT mass scale $\L_G \sim 10^{15}$~GeV. 
This is the scenario studied in \cite{bp96},
 with a baryogenesis temperature $T_B \sim M_1 \sim 10^{10}$~GeV.
Such a high temperature has important implications for the dark matter problem
due to the abundance of thermally produced gravitinos.    

In summary, decays of heavy Majorana neutrinos provide a simple and elegant
explanation of the origin of matter in our universe. Before recombination
and primordial nucleosynthesis, baryogenesis would then be the next important
period in the early universe, at a temperature $T_B \sim 10^{10}$~GeV
and a corresponding time $t_B \sim 10^{-26}$~s. An unequivocal prediction of
this picture is that neutrinos
are Majorana particles with masses below $0.2$~eV, which will be tested by
forthcoming laboratory experiments and by cosmology.\\

\noindent
\textbf{Acknowledgment}\\
\noindent
M.P.~was supported by the EU network ``Supersymmetry and the
Early Universe'' under contract no.~HPRN-CT-2000-00152.


\end{document}